\pdfoutput=1

\documentclass[11pt]{article}

\usepackage[]{acl}
\usepackage{times}
\usepackage{latexsym}
\usepackage{booktabs}
\usepackage{amsmath, amssymb, graphicx, xcolor, multirow, multicol, comment, subcaption, url, float, etoolbox, adjustbox, pgf, soul, geometry, colortbl, booktabs}
\usepackage{amsmath}
\usepackage[ruled,vlined]{algorithm2e} 
\usepackage[T1]{fontenc}

\usepackage[utf8]{inputenc}

\usepackage{microtype}

\usepackage{inconsolata}

\usepackage{graphicx}
\usepackage{xcolor}

\usepackage{microtype}
\definecolor{lightred}{rgb}{1.0, 0.8, 0.8}
\definecolor{lightblue}{rgb}{0.8, 0.9, 1.0}
\definecolor{lightgreen}{rgb}{0.8, 1.0, 0.8}
\definecolor{lightyellow}{rgb}{1.0, 1.0, 0.8}
\definecolor{lightpurple}{rgb}{0.9, 0.8, 1.0}
\definecolor{lightorange}{rgb}{1.0, 0.9, 0.8}

\title{Investigating Prosodic Signatures via Speech Pre-Trained Models for Audio Deepfake Source Attribution}

\author{
  Orchid Chetia Phukan$^{1*}$, 
  Drishti Singh$^{1}\thanks{\footnotesize{Equal Contribution as first authors}}$,
  Swarup Ranjan Behera$^{2}$\\
  \textbf{Arun Balaji Buduru}$^{1}$,
  \textbf{Rajesh Sharma}$^{1,3}$\\
  \textsuperscript{1}IIIT-Delhi, India, 
  \textsuperscript{2}Independent Researcher, India,
  \textsuperscript{3}University of Tartu, Estonia\\ 
  \texttt{\textbf{Correspondence:} \textcolor{blue}{orchidp@iiitd.ac.in}}
}

\begin{document}
\maketitle
\begin{abstract}
In this work, we investigate various state-of-the-art (SOTA) speech pre-trained models (PTMs) for their capability to capture prosodic signatures of the generative sources for audio deepfake source attribution (ADSD). These prosodic characteristics can be considered one of major signatures for ADSD, which is unique to each source. So better is the PTM at capturing prosodic signs better the ADSD performance. We consider various SOTA PTMs that have shown top performance in different prosodic tasks for our experiments on benchmark datasets, ASVSpoof 2019 and CFAD. x-vector (speaker recognition PTM) attains the highest performance in comparison to all the PTMs considered despite consisting lowest model parameters. This higher performance can be due to its speaker recognition pre-training that enables it for capturing unique prosodic characteristics of the sources in a better way. Further, motivated from tasks such as audio deepfake detection and speech recognition, where fusion of PTMs representations lead to improved performance, we explore the same and propose \textbf{FINDER} for effective fusion of such representations. 
With fusion of Whisper and x-vector representations through \textbf{FINDER}, we achieved the topmost performance in comparison to all the individual PTMs as well as baseline fusion techniques and attaining SOTA performance.

\end{abstract}

\vspace{-0.2cm}
\section{Introduction}

Imagine waking up to find your voice used in a viral audio clip, falsely implicating you in a scandal. This increasingly plausible scenario highlights the growing threat of audio deepfakes. With advancements in text-to-speech (TTS) and voice conversion (VC) technologies, malicious actors can now create synthetic audio that is nearly indistinguishable from authentic recordings. From high-profile frauds~\cite{stupp2019fraudsters} to the viral spread of falsified audio targeting political figures~\cite{BBC}, the misuse of synthetic audio for financial scams and misinformation underscores the urgent need for reliable detection methods. As the authenticity of audio content becomes increasingly difficult to verify, the importance of audio deepfake detection (ADD) has never been more pressing.

Despite advancements in ADD, research has primarily focused on binary classification - distinguishing real from fake audio~\cite{ref12,ref13,ref14,ref15,ref16,ref18,ref19,ref62,ref63,ref64,ref65}. This approach, while effective in its simplicity, lacks the granularity needed to address a crucial aspect of deepfake detection: source attribution. Audio deepfake source attribution (ADSD) goes beyond merely identifying whether audio is real or fake; it seeks to uncover the specific tool or model responsible for generating the synthetic audio~\cite{SA1,SA2,SA3}. This capability is vital for improving the explainability of detection systems and enabling targeted countermeasures, especially in high-stakes contexts such as audio forensics and intellectual property protection. 

Generative sources such as TTS, VC, etc. systems embed their unique prosodic characteristics, such as pitch, tone, rhythm, and intonation, into their generated audios, reflecting the inherent design and processing patterns of the generative system. These prosodic signatures are vital for accurately identifying the source and can be considered one of the major fingerprint. In this study, we investigate various state-of-the-art (SOTA) speech pre-trained models (PTMs) for capturing these prosodic signatures of source for ADSD as such PTMs has shown significant potential in advancing ADSD \cite{klein24_interspeech}. We consider PTMs that have shown efficacy in various prosodic tasks such as speech emotion recognition (SER), depression detection, etc.  Motivated by tasks like ADD \cite{chetia-phukan-etal-2024-heterogeneity} and speech recognition \cite{arunkumar22b_interspeech}, we also investigate the fusion of different PTM representations and propose, \textbf{FINDER} (\textbf{F}us\textbf{I}on through Re\textbf{N}yi \textbf{D}iv\textbf{ER}gence) for effective fusion. We believe, we are the first work, to the best of our knowledge, for exploring fusion of PTMs representations for ADSD. 

\noindent \textbf{To summarize, the main contributions}:
\begin{itemize}
    \item We give a comprehensive comparative study of SOTA speech PTMs for investigating their capacity of capturing prosodic signatures for ADSD.
    \item We show that x-vector, a speaker recognition PTM, achieves the highest performance and this behavior can be attributed to its speaker recognition pre-training that enables it to capture prosodic features better.
    \item We propose \textbf{FINDER}, a novel framework that leverages renyi divergence as a fusion mechanism for fusion of PTMs representations. 
\end{itemize}


\section{Related Work}
Early work on ADSD introduced the problem of identifying attacker signatures, showing that representations from RNN can characterize both seen and unseen attackers with high accuracy \cite{SA5}. Subsequent studies focused on detecting vocoder-specific fingerprints, revealing that vocoders leave identifiable traces in generated audio \cite{SA1, SA6}. Building on this, methods such as t-SNE visualization and ResNet-based architectures further improved fingerprint detection accuracy \cite{SA6}. \citealt{SA4} proposed VFD-Net, a patch-wise supervised contrastive learning method, which achieved robust performance under cross-set and audio compression conditions. More recent work \cite{SA7} has shown potential of using PTMs as backbones for improved ADSD. In this study, for the first time, we investigate SOTA PTMs for assessing their capability of capturing unique prosodic signatures of sources for better ADSD. 




    




    

\section{Pre-Trained Models}
 Wav2vec2 \cite{baevski2020wav2vec} and WavLM \cite{chen2022wavlm} are monolingual PTMs. Wav2vec2 trained on the LibriSpeech dataset, masks the input in latent space, it has shown effectiveness in prosodic tasks such as SER \cite{pepino21_interspeech}. WavLM showed SOTA performance on SUPERB including various prosodic tasks. XLS-R \cite{babu22_interspeech} and Whisper \cite{radford2023robust} are multilingual PTMs. XLS-R was pre-trained on 128 languages for 436k hours of unlabeled speech while Whisper \cite{radford2023robust} on 96 languages for 680k hours of labelled data. XLS-R shows good performance in ML-SUPERB \cite{shi23g_interspeech} that includes prosodic tasks while Whisper shows potential for SER \cite{feng2023peft}. In addition to these PTMs, we consider, x-vector \cite{8461375}, trained for speaker recognition. It excels in various prosodic tasks such as SER \cite{chetiaphukan23_interspeech}, shout intensity prediction \cite{fukumori2023investigating}, depression detection \cite{9746068}, and so on. We also consider, Wav2Vec2-emo\footnote{\url{https://huggingface.co/speechbrain/emotion-recognition-wav2vec2-IEMOCAP}} fine-tuned for SER, as SER is a prosodic task and we think its representations might be helpful for ADSD. 
Additional details regarding the above PTMs are provided in Appendix \ref{ptms}. 

\section{Modeling}

We consider two downstream networks i.e. fully connected network (FCN) and CNN with individual PTM representations. The FCN model consists of 3 dense layers with CNN model has four convolutional layers followed by three dense layers with 256, 128, and 64 neurons followed by the output layer. For CNN, we use two convolution blocks each consisting of 1D-CNN and max-pooling layer followed by flattening and FCN with similar configuration as FCN network above. 
Hyperparameters detail is given in Appendix \ref{hyper}. \par

\begin{figure}[bt] 
    \centering
    \includegraphics[width=0.475\textwidth]{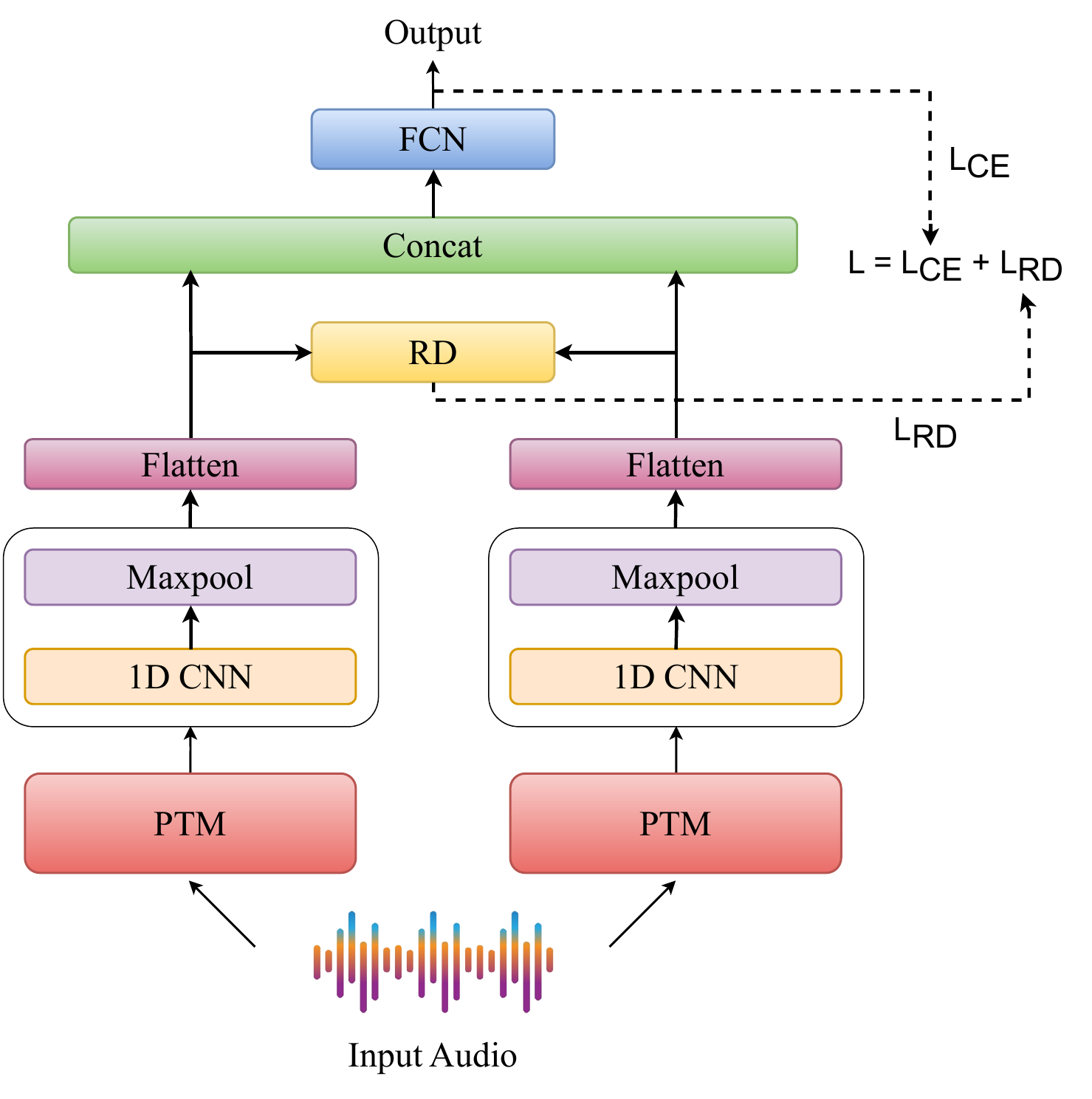} 
    \caption{Proposed Framework \textbf{FINDER}: RD and FCN stand for renyi divergence and fully connected network, respectively; $L$, $L_{CE}$, and $L_{RD}$ represent the total loss, cross-entropy loss, and renyi divergence loss, respectively.}
    \label{fig:archi_rd}
\end{figure}

\noindent\textbf{FINDER}: We propose \textbf{FINDER} for effective fusion of PTMs representations. The architecture is given in Figure \ref{fig:archi_rd}. The PTMs representations are passed through two convolution blocks with same configuration as CNN model built for modeling with individual PTM representations above.

Features are flattened after the convolution blocks and linearly projected to 120-dimensional size to keep the same dimensions and also for computational constraints. The projected features then passed through the renyi divergence (RD). RD is a measure of divergence or dissimilarity between two probability distributions \cite{van2014renyi}. Here, we frame RD as a loss function that calculates the divergence between the feature representations of two different PTMs. We aim to reduce the disimilarity between the feature representations and make it closer to each other. ${e}_{\text{a}}$ and ${e}_{\text{b}}$ be the feature space for two PTMs networks.

RD between the two feature distributions ${e}_{\text{a}}$ and ${e}_{\text{b}}$ is given by:

\[
\mathcal{L}_{RD} = \frac{1}{\alpha - 1} \log \left( \sum_{i=1}^D (e_{\text{a},i} + \epsilon)^\alpha (e_{\text{b},i} + \epsilon)^{1 - \alpha} \right)
\]
where \( D \) is the embedding dimension, \( \alpha > 1 \) controls the order of the divergence, and \( \epsilon \) is a small constant for numerical stability.

Finally, we add the RD loss ${L}_\text{RD}$ to the cross-entropy loss ${L}_\text{CE}$ for joint optimization. Total loss is given as:
\[
    \mathcal{L} = \lambda \mathcal{L}_{CE} + (1-\lambda) \mathcal{L}_{RD}
\]
where $\lambda$ is a hyperparameter and weightage parameter for the losses. 

\section{Experiments}
\textbf{Benchmark Datasets}: We use two benchmark datasets for our experiments namely ASVSpoof 2019 (ASV) \cite{wang2020asvspoof} and FAD Chinese Dataset (CFAD) \cite{ma2024cfad}. We combine the train, validation and testing splits for ASV and in resultant we got 19 classes (A01 to A19) as fake audio sources, while CFAD has 9 classes (0 - 8) for source classes. We followed 5-fold cross validation for ASV and followed official split given for CFAD. For more details on the datasets and data preprocessing, refer to Appendix \ref{dataaaappendix}. \par

\noindent \textbf{Training Details}: We use softmax as the activation function in the output layer of all the models that outputs class probabilites. We use Adam as optimizer with learning rate of \(10^{-3}\). We trained all the models for 40 epochs with a batch size of 32. We use cross-entropy as the loss function for all the models. We use early stopping and dropout for preventing overfitting. For experiments with \textbf{FINDER}, we set \( \alpha = 2 \), \( \epsilon = 0.1\), and \(\lambda = 0.4\) and keep these values constant throughout the experiments as we got better results with these values through some preliminary exploration. 



\begin{table*}[bt]
\centering
\scriptsize
\caption{Performance Comparison of individual PTMs representations on ASV and CFAD; All the scores are average of 5-folds and in \%; High Accuracy, Low EER better the model}
\renewcommand{\arraystretch}{0.9} 
\begin{tabular}{lcccc|cccc}
\toprule
\textbf{Representations} & \multicolumn{4}{c|}{\textbf{ASV}} & \multicolumn{4}{c}{\textbf{CFAD}} \\
\cmidrule(r){2-5} \cmidrule(l){6-9}
 & \multicolumn{2}{c}{\textbf{FCN}} & \multicolumn{2}{c|}{\textbf{CNN}} & \multicolumn{2}{c}{\textbf{FCN}} & \multicolumn{2}{c}{\textbf{CNN}} \\
\cmidrule(r){2-3} \cmidrule(r){4-5} \cmidrule(r){6-7} \cmidrule(l){8-9}
 & \textbf{Accuracy} & \textbf{EER} & \textbf{Accuracy} & \textbf{EER} & \textbf{Accuracy} & \textbf{EER} & \textbf{Accuracy} & \textbf{EER} \\
\midrule
\textbf{Wav2vec2} & 45.25 & 21.54 & 61.75 & 7.76 & 49.25 & 29.20 & 74.50 & 10.20 \\
\textbf{WavLM} & 33.48 & 15.25 & 45.46 & 10.50 & 32.23 & 22.23 & 35.78 & 27.60 \\
\textbf{XLS-R} & 63.98 & 11.55 & 79.04 & 4.01 & 50.25 & 19.30 & 76.90 & 9.50 \\
\textbf{Whisper} & 75.69 & 9.85 & 87.03 & 4.01 & 70.14 & 15.85 & 85.01 & 8.10 \\
\textbf{x-vector} & \textbf{87.48} & \textbf{4.42} & \textbf{97.60} & \textbf{2.03} & \textbf{74.58} & \textbf{10.02} & \textbf{91.39} & \textbf{4.40} \\
\textbf{Wav2vec2-emo} & 78.45 & 8.40 & 86.35 & 2.50 & 65.30 & 12.12 & 86.50 & 8.60 \\
\bottomrule
\end{tabular}
\label{tab:single_ptms_table}
\end{table*}

\begin{table*}[bt]
\centering
\scriptsize
\caption{Performance Comparison of Fusion Methods on ASV and CFAD; All the scores are average of 5-folds and in \%; High Accuracy, Low EER better the model}
\renewcommand{\arraystretch}{1.0} 
\begin{tabular}{lcccc|cccccccc} 
\toprule
\textbf{Representations} & \multicolumn{4}{c|}{\textbf{ASV}} & \multicolumn{4}{c}{\textbf{CFAD}} \\ 
\cmidrule(r){2-5} \cmidrule(lr){6-9} 
 & \multicolumn{2}{c}{\textbf{Concatenation}} & \multicolumn{2}{c|}{\textbf{FINDER}} & \multicolumn{2}{c}{\textbf{Concatenation}} & \multicolumn{2}{c}{\textbf{FINDER}} \\
\cmidrule(r){2-3} \cmidrule(r){4-5} \cmidrule(r){6-7} \cmidrule(lr){8-9}
 & \textbf{Accuracy} & \textbf{EER} & \textbf{Accuracy} & \textbf{EER} & \textbf{Accuracy} & \textbf{EER} & \textbf{Accuracy} & \textbf{EER} \\
\midrule
\textbf{Wav2vec2 + Wav2vec2-emo} & 96.57 & 0.67	& 96.69 & 0.63 & 89.80	 & 3.60 & 93.32	 & 3.51\\
\textbf{WavLM + Wav2vec2-emo} & 93.60 & 1.10	& 94.60 & 1.08 & 85.65 & 5.70 & 88.03 & 5.55\\
\textbf{XLS-R + Wav2vec2-emo} & 91.23 & 1.10 & 94.35 & 1.04 & 92.47 & 2.79 & 95.72& 2.64\\
\textbf{Whisper + Wav2vec2-emo} & 96.03 & 0.66 & 96.66 & 0.63 & 94.95	 & 1.50	 & 98.47	 & 1.41\\
\textbf{x-vector + Wav2vec2-emo} & 93.12	 & 1.33 & 98.62	 & 0.37 & 91.79	 & 2.80	 & 96.71	 & 2.50\\
\textbf{x-vector + Wav2vec2} & 96.25	 & 0.63	 & 98.16	 & 0.33	 &  85.50	 & 5.40 & 86.77	 & 4.40\\
\textbf{WavLM + Wav2vec2} & 96.38	 & 0.69	 & 96.99	 & 0.65	 & 89.75	 & 2.84	 & 96.26	 & 2.61\\
\textbf{Whisper + Wav2vec2} & 96.87	 & 0.51 & 97.21 & 0.50 & 94.11	 & 2.04	 & 97.59	 & 1.91\\
\textbf{XLS-R + Wav2vec2} & 96.09	 & 0.65	 & 96.54 & 0.57	 & 94.93	 & 1.45	 & 98.09	 & 1.24\\
\textbf{WavLM + XLS-R} & 86.92	 & 2.30	 & 87.40	 & 2.04	 & 92.63	 & 1.80	 & 98.75	 & 1.17\\
\textbf{Whisper + XLS-R} & 93.99	 & 1.10	 & 94.50	 & 1.01	 & 89.50 & 3.40 & 94.98 & 2.97\\
\textbf{x-vector + XLS-R} & 96.58	 & 1.20	& 97.84	 & 0.35 & 90.87	 & 3.30	 & 95.58	 & 1.60\\
\textbf{Whisper + WavLM} & 94.26	 & 0.85	 & 94.69	 & 0.84	 & 88.86	 & 3.84	 & 94.28	 & 3.08\\
\textbf{x-vector + WavLM} & 95.85	 & 0.40 & 98.37	 & 0.36	 & 64.32	 & 10.10 & 78.77	 & 8.17	\\
\textbf{Whisper + x-vector} & \textbf{97.16}	 & \textbf{0.32} & \textbf{98.91}	 & \textbf{0.26}		 & \textbf{95.00} & \textbf{1.10} & \textbf{99.01} & \textbf{1.07}  \\
\bottomrule
\end{tabular}
\label{tab:combination_ptms_table}
\end{table*}

\noindent \textbf{Experimental Results}: We use accuracy and equal error rate (EER) as the evaluation metrics for experiments as used by previous works on ADSD \cite{klein24_interspeech} and ADD \cite{ref15}. For EER, we present the average scores of one-vs-all.

Table \ref{tab:single_ptms_table} presents the results of downstream models trained on individual PTM representations. x-vector consistently delivers best results across both the datasets, achieving high accuracy and lower EER. This performance can be traced back to its speaker recognition pre-training that equips x-vector to better capture prosodic features also consistent across various prosodic tasks \cite{chetiaphukan23_interspeech, fukumori2023investigating, 9746068}. In contrast, monolingual PTMs like Wav2vec2 and WavLM underperformed due to their limited capacity to capture source-specific prosodic features. The performance of Wav2vec2-emo is a known behavior as it was trained for SER, but not better than x-vector. Whisper and XLS-R also shows better performance than monolingual PTMs as seen in previous research for ADD \cite{chetia-phukan-etal-2024-heterogeneity} that multilingual PTMs capture diverse pitches, tones, etc. prosodic characteristics better than monolingual PTMs. We also plots the t-SNE plots of the PTMs representations in Appendix Figure \ref{tsnecfad} and \ref{tsneasv}. The plots adds to the performance of x-vector as we can observe better clusters across the source classes. Also, CNN models consistently outperform FCN models. 

Table \ref{tab:combination_ptms_table} presents the results for fusion of PTMs representations through baseline concatentation fusion technique and \textbf{FINDER}. For the baseline, we use the same modeling paradigm except the RD loss. We observe that results of fusion of PTMs representations through both the fusion techniques are better than the individual PTM representations, thus, showing their complementary nature. Fusion of PTM representations through \textbf{FINDER} consistently beat the concatenation based baseline fusion techniques showing its effectiveness. Fusion of Whisper and x-vector through \textbf{FINDER} shows the best performance across both the datasets.

\begin{table}[htbp]
\scriptsize
\centering
\begin{tabular}{l|l|l|c}
\toprule
\textbf{Dataset} & \textbf{Model} & \textbf{Accuracy(\%)} & \textbf{EER(\%)}\\
\midrule
\textbf{ASV} & \textbf{FINDER (Whisper + x-vector)} & \textbf{98.91} & \textbf{0.26} \\
    & MiO(Whisper + x-vector) & 97.75 & 0.68\\
    & AASIST(Wav2vec2) & 63.81 & 5.68\\
\midrule
\textbf{CFAD} & \textbf{FINDER} & \textbf{99.01}  & \textbf{1.07} \\
    & MiO(Whisper + x-vector) & 97.31 &  2.15 \\
    & AASIST(Wav2vec2) & 77.92 & 9.69\\
\bottomrule
\end{tabular}
\caption{Comparison to previous SOTA works}
\label{eer_comparison_table}
\end{table}

\noindent \textbf{Comparison to Previous Works}: As we have considered all the source classes across train, validation, and test split for ASV, so we can't directly compare our results to previous works. For CFAD, we are the first one to perform ADSD, so there is not previous work to compare to. So, we reimplemented some of the SOTA methods for ADSD and ADD and compared it our results. For ADD, we consider MiO \cite{chetia-phukan-etal-2024-heterogeneity}, an SOTA method that proposed combination of PTM representations. We implemented with Whisper and x-vector representations i.e. the best performing pair. For ADSD, we implemeted AASIST \cite{jung2022aasist} as downstream with Wav2vec2 representations used by \citealt{klein24_interspeech}. Table \ref{eer_comparison_table} presents the comparison of the proposed with the SOTA methods. We observe the \textbf{FINDER} outperforms both the methods and attains SOTA performance showing its effectiveness for ADSD.


\vspace{-0.3cm}
\section{Conclusion}  
In this work, we investigate various SOTA speech PTMs for their ability to capture prosodic signatures of generative sources for ADSD. We evaluate monolingual, multilingual, and speaker recognition PTMs on benchmark datasets (ASV, CFAD), finding that x-vector, a speaker recognition PTM, outperforms others due to its ability to capture better source-specific prosody. Further, we explored fusion of PTM representations for ADSD and propose \textbf{FINDER} for the same. With fusion of x-vector and Whisper representations through \textbf{FINDER}, we achieve the topmost performance surpassing both individual PTMs and baseline fusion techniques and attains SOTA performance.

\section{Limitations}
One major limitation of our work is the proposed systems are not built for open-vocabulary ADSD. It can only identify fake audios that are generated by the generative systems present in the datasets, considered in our study. In our future work, we will work towards building systems for open-vocabulary ADSD.

Another limitation is the experimentation with limited downstream networks, previous research has shown that the downstream performance changes with the downstream modeling technique \cite{zaiem23b_interspeech}. Here, we have experimented with only FCN and CNN. In future, we will explore further varied downstream networks for ADSD.

\section{Ethics Statement}
ASV and CFAD are publicly available and widely recognized benchmarks for audio deepfake research and have been automatically identified to protect speaker privacy. Our proposed method aims to enhance ADSD, supporting efforts to combat misinformation, fraud, and misuse of generative models. We acknowledge the potential risks associated with generative technologies and emphasize that our work is solely intended for improving detection systems and promoting cybersecurity.

\bibliography{acllatex}

\appendix

\section{Appendix}
\label{sec:appendix}
\subsection{Detailed Information of PTMs}
\label{ptms}
    In this section, we present detailed information regarding the PTMs utilized in our study.
\begin{itemize}
    \item \textbf{Wav2vec2\footnote{\url{https://huggingface.co/facebook/wav2vec2-base}}}: It is trained in self-supervised manner to perform a contrastive task based on the quantization of jointly learned latent representations. It is trained on 960 hrs of audio LibriSpeech data, particularly English language. We are using the Wav2Vec2-base model with approximately 95 million parameters. 

    \item \textbf{Wav2vec2-emo}: It is a fine-tuned version of Wav2vec2 for SER on the benchmark IEMOCAP dataset using the general purpose speech toolkit - SpeechBrain. Similar to Wav2Vec 2.0, it has 95.04 million parameters. 

    \item \textbf{XLS-R\footnote{\url{https://huggingface.co/facebook/wav2vec2-xls-r-300m}}}: It is a cross-lingual learnng model trained in self-supervised manner, based on the Wav2Vec2 framework. The model is trained on approximately 500k hrs of open source speech audio data, spanning over 128 languages. For our experiments, we are utilizing the version with 300 million parameters.

    \item \textbf{Whisper\footnote{\url{https://huggingface.co/openai/whisper-base}}}: It is a model trained in a multitask manner on internet data of about 680k hours, consisting of multilingual and multitask supervision. Whisper demonstrates ability to generalize on diverse datasets and domains without the need for fine-tuning, in a zero-shot setting. Whisper shows improved performance on speech recognition over XLS-R. We have used the base model with 74 million parameters.

    \item \textbf{WavLM\footnote{\url{https://huggingface.co/microsoft/wavlm-base}} }: It is self-supervised PTM designed to address the challenges of learning universal speech representations for diverse speech processing tasks. WavLM uses masked speech prediction with speech denoising during pre-training, enabling it to model both spoken content and non-ASR tasks effectively. It is trained on 960k hours of Librispeech English data, outperforming models like Wav2vec2 and HuBERT. We have used the base version with 94.70 million parameters. 

    \item \textbf{X-vector\footnote{\url{https://huggingface.co/speechbrain/spkrec-xvect-voxceleb}}}: It is a time-delay neural network (TDNN) trained in a supervised manner for speaker recognition. It is trained on Voxceleb 1+ Voxceleb2 training data, using the general purpose speech toolkit - Speechbrain. It has achieved SOTA performance in speaker recognition, outperforming models like i-vector. We are using the Speechbrain model with approx 4.2 million parameters. 
\end{itemize}

We resample the audios to 16 KHz before passing it to the PTMs and extract representations from the last hidden state of the PTMs by average pooling. We 

\subsection{Benchmark Datasets}
\label{dataaaappendix}
\noindent \textbf{ASV}: It was was developed to advance research in detecting audio deepfake detection and protecting automatic speaker verification systems from manipulation. It encompasses three major spoofing types: synthetic speech, converted speech, and replay attacks, generated using SOTA neural acoustic and waveform models. The dataset comprises bonafide (genuine) and spoofed audio samples, sourced from 19 synthetic systems, with recordings at a 16 kHz sample rate and 16-bit depth. Bonafide recordings include diverse speakers, capturing a range of accents, speaking styles, and vocal characteristics. Spoofed samples are predominantly produced using SOTA voice conversion (VC) and text-to-speech (TTS) methods, ensuring high clarity and naturalness. For this study, we leverage the 19 spoofed classes for neural generator attribution. Statistics are presented in Table \ref{asvdata}.

\begin{table}[bt]
\centering
\scriptsize
\begin{tabular}{lrr}
\toprule
\textbf{Category}            & \textbf{Frequency} \\
\midrule
Dev                          & 24,844             \\
Train                        & 25,380             \\
Eval                         & 71,237             \\
\midrule
\textbf{Total Samples}       & \textbf{121,461}   \\
\midrule
Real                   & 12,483             \\
Total Fake Audio Samples    & 108,978            \\
\bottomrule
\end{tabular}
\caption{ASV Statistics}
\label{asvdata}
\end{table}

\noindent \textbf{CFAD}: It addresses the lack of public Chinese datasets under noisy conditions for fake audio detection. It includes bonafide and fake audio generated by 12 advanced speech generation techniques. To simulate real-world conditions, three noise datasets were added at five different signal-to-noise ratios (SNRs).  Spoofed samples were synthesized using 11 SOTA TTS and VC methods, as well as neural network-based speech generation models. The dataset facilitates evaluation using metrics such as Equal Error Rate (EER) and Tandem Detection Cost Function (t-DCF). CFAD contributes significantly to audio forensics, enabling the identification of manipulated content and the attribution of spoofing algorithms.

\noindent \textbf{Data Pre-Processing}: We resample the audios to 16 KHz before passing it to the PTMs and extract representations from the last hidden state of the PTMs by average pooling. We extract representations of 768-dimensions for Wav2vec2, Wav2vec2-emo and WavLM. We get representations of 512 for Whisper-encoder and x-vector and 1024 for XLS-R.

\subsection{Hyperparameters and System Configurations}
\label{hyper}
The first layer of convolution block of the CNN model has 256 filters and a kernel size of 3, followed by batch normalization, and max pooling (pool size 2). The second layer uses 128 filters and a kernel size of 3, succeeded by followed by batch normalization, and max pooling (pool size 2. The trainable parameters for CNN models with individual PTMs representations ranged from 0.8 to 1.2 million parameters depending on the input representation dimension size. Also, for the fusion experiments, the trainable parameters of the models range from 1.3 to 1.5 million parameters. 

We use \textit{Tensorflow} library for carrying out our experiments. We use A5000 GPU  for running our experiments. The models and codes curated for this study will be released after the double-blind reviewing process.

\begin{table}[bt]
\centering
\scriptsize
\begin{tabular}{lrr}
\toprule
\textbf{Category}         & \textbf{Total} & \textbf{Fake} \\
\midrule
Train                      & 138,400        & 25,600         \\
Validation                 & 14,400         & 9,600          \\
Test                       & 42,000         & 28,000         \\
\midrule
\textbf{Total Samples}    & \textbf{194,800} & \textbf{63,200} \\
\bottomrule
\end{tabular}
\caption{CFAD Statistics}
\label{tab:fad2_dataset}
\end{table}

\begin{figure*}[htbp]
    \centering
    \subfloat[WavLM]{\includegraphics[width=0.425\textwidth]{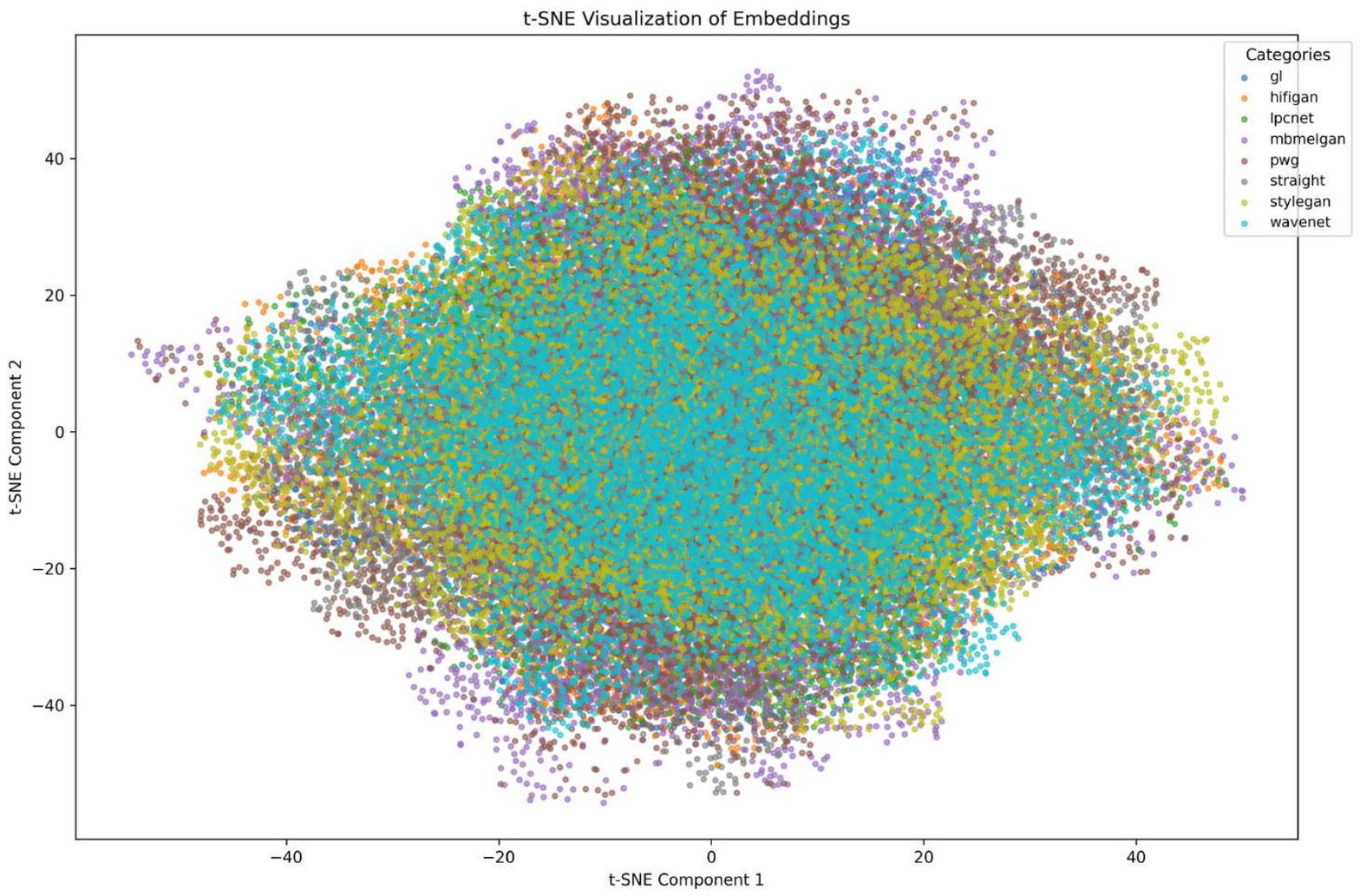}} 
    \subfloat[Whisper]{\includegraphics[width=0.425\textwidth]{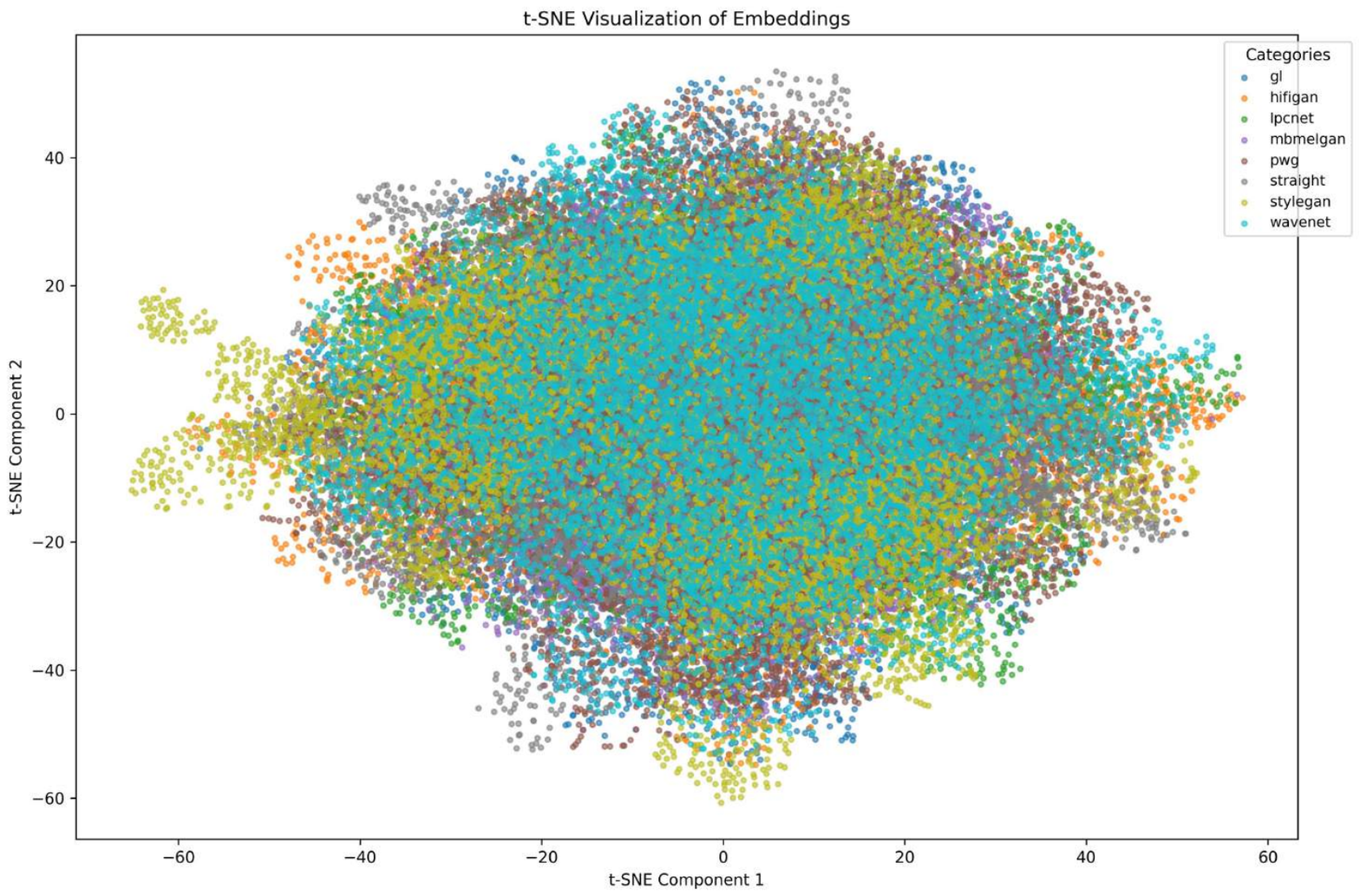}} \\

    \vspace{0.5cm}  
    \subfloat[x-vector]{\includegraphics[width=0.425\textwidth]{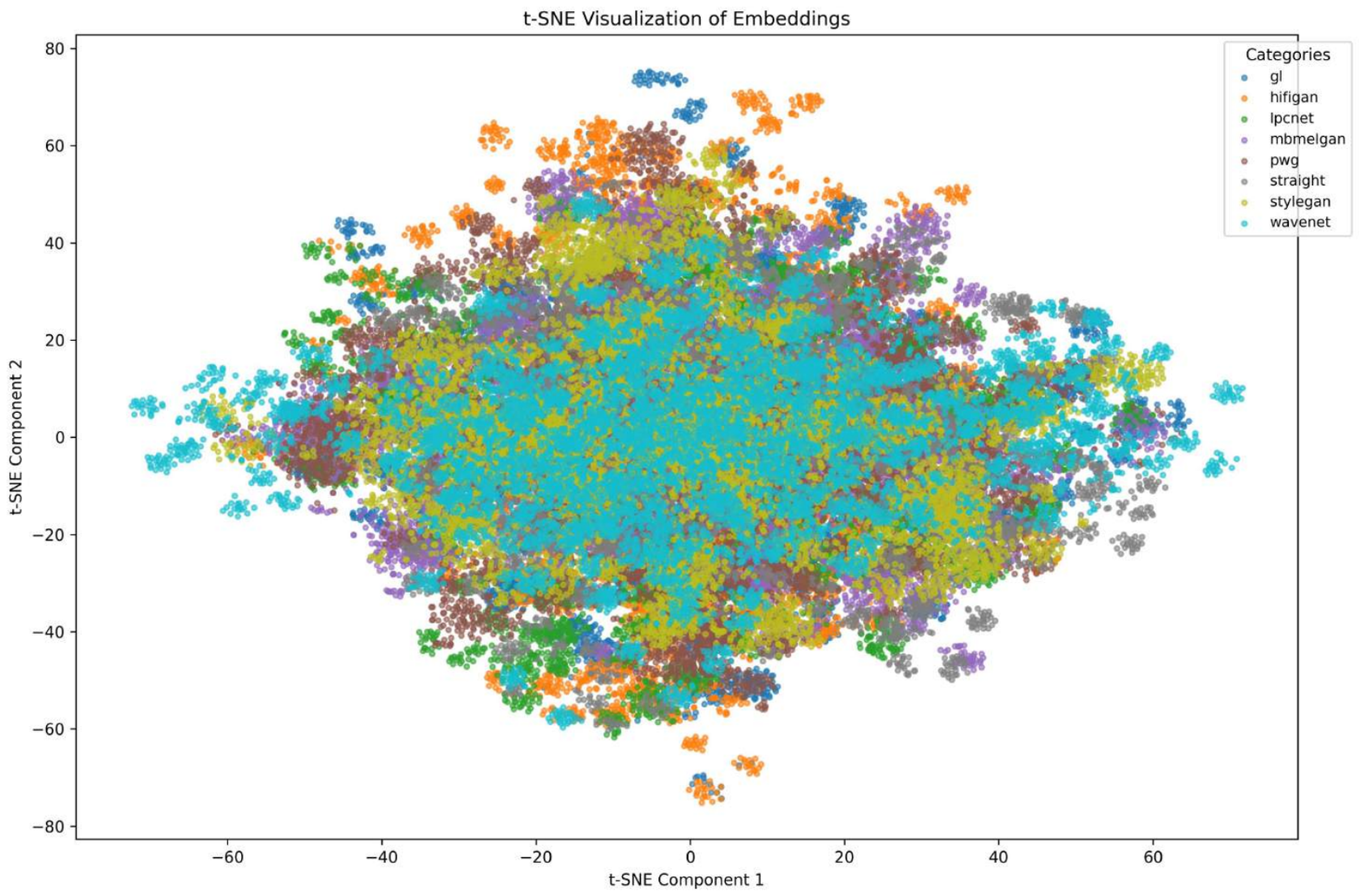}} 
    \subfloat[Wav2vec2-emo]{\includegraphics[width=0.425\textwidth]{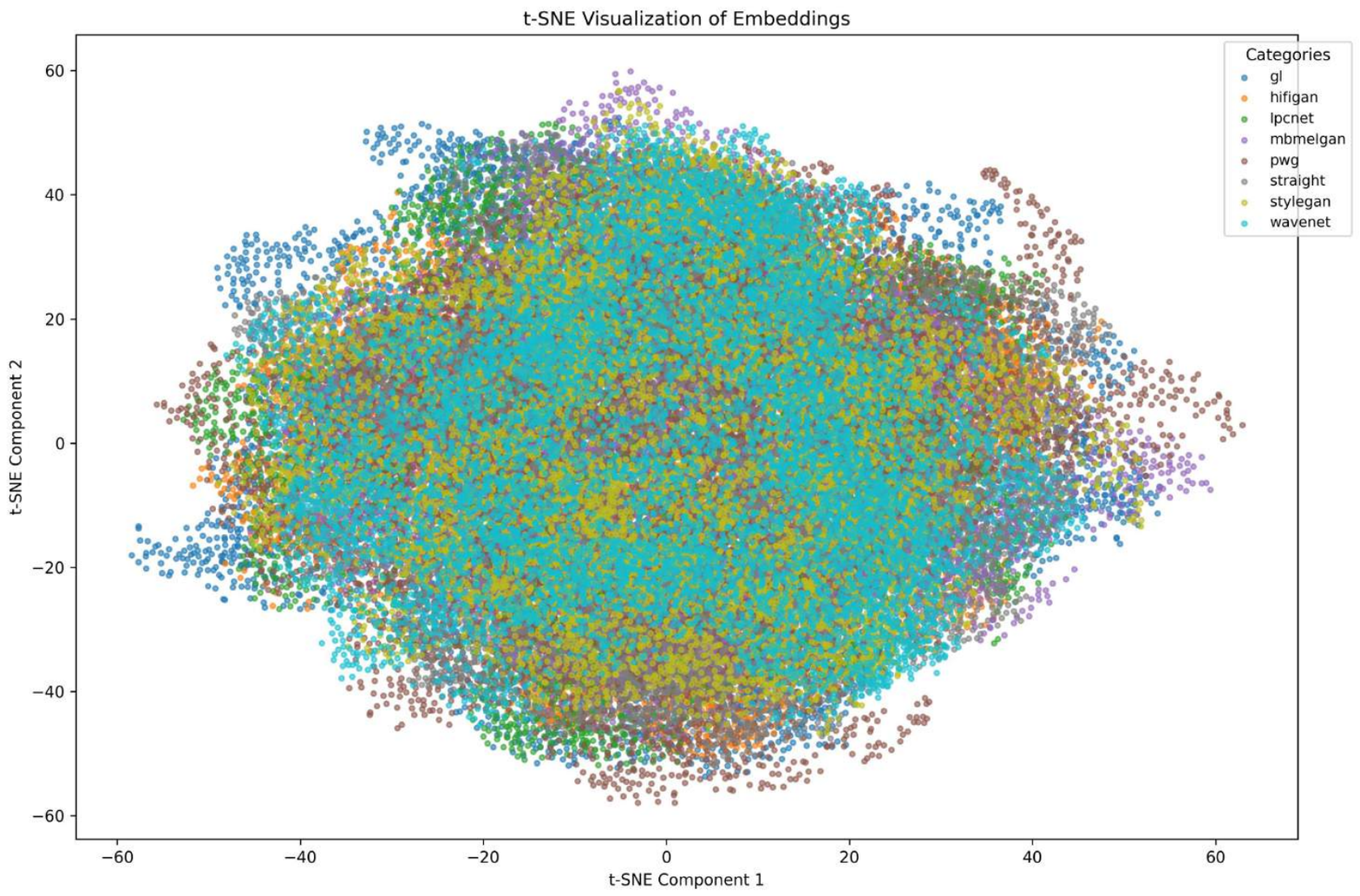}} 

    \caption{Representation Space Visualization of PTMs for CFAD}
    \label{tsnecfad}
\end{figure*}

\begin{figure*}[htbp]
    \centering
    \subfloat[WavLM]{\includegraphics[width=0.425\textwidth]{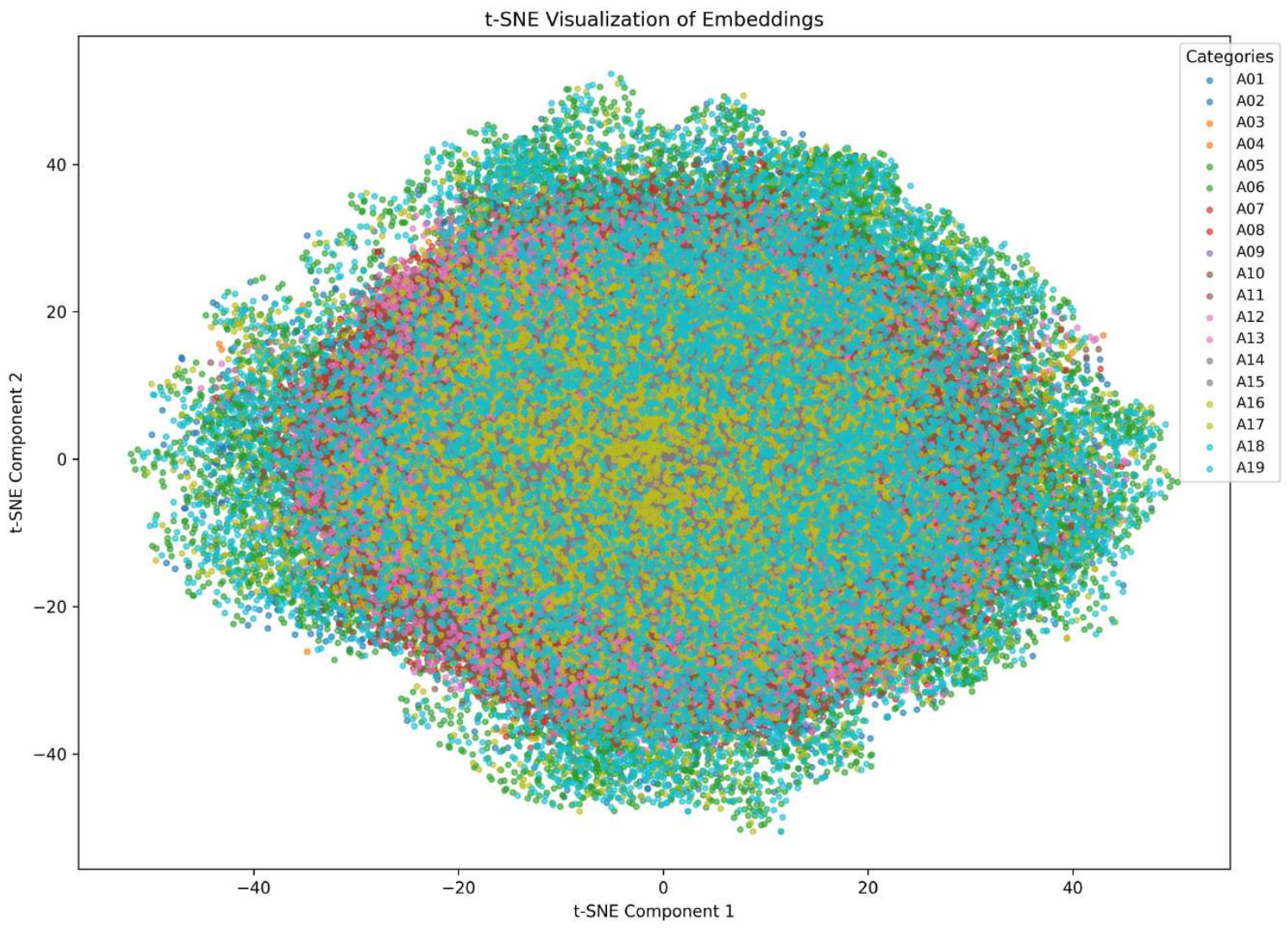}} 
    \subfloat[Whisper]{\includegraphics[width=0.425\textwidth]{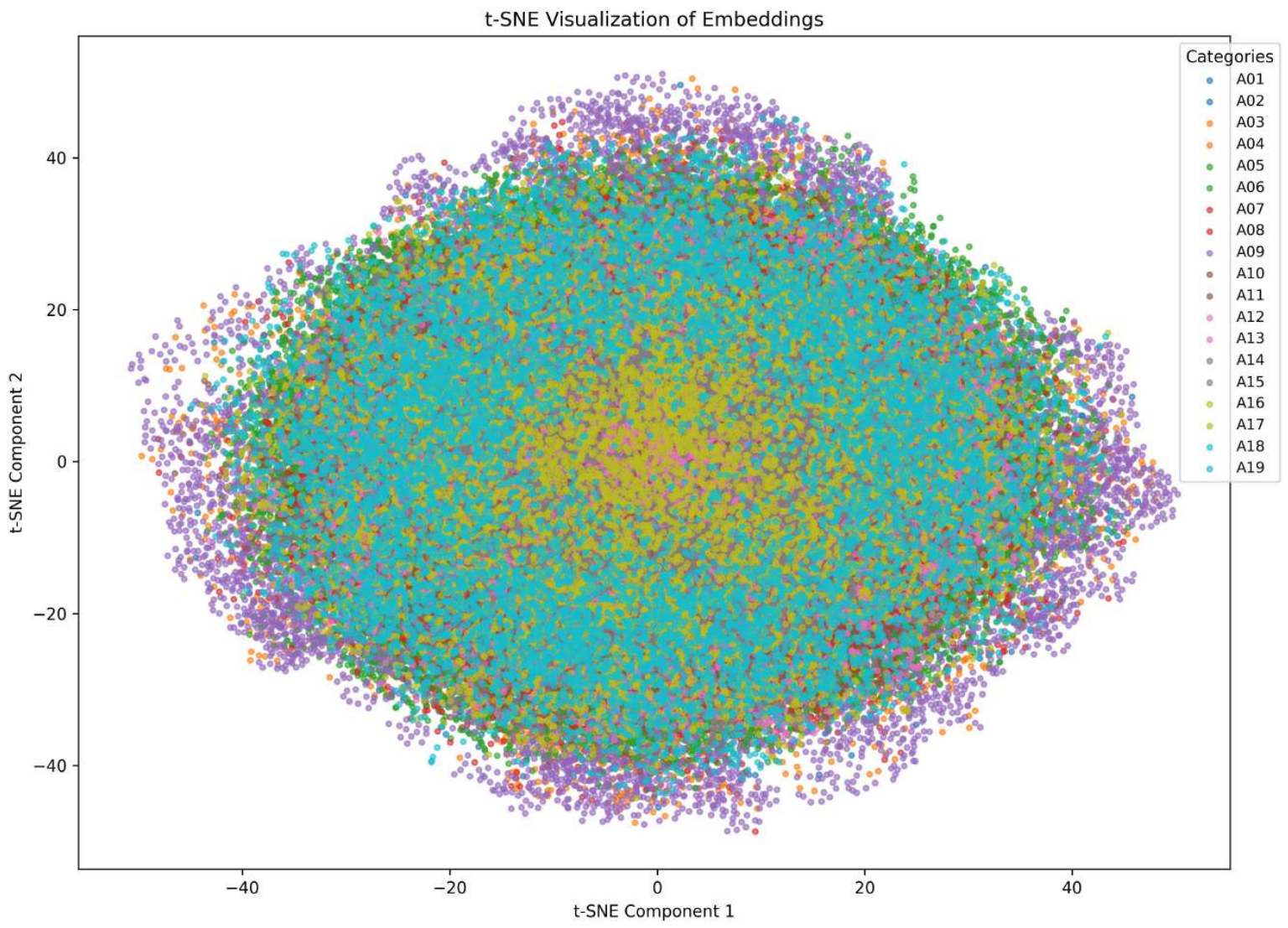}} \\

    \subfloat[x-vector]{\includegraphics[width=0.425\textwidth]{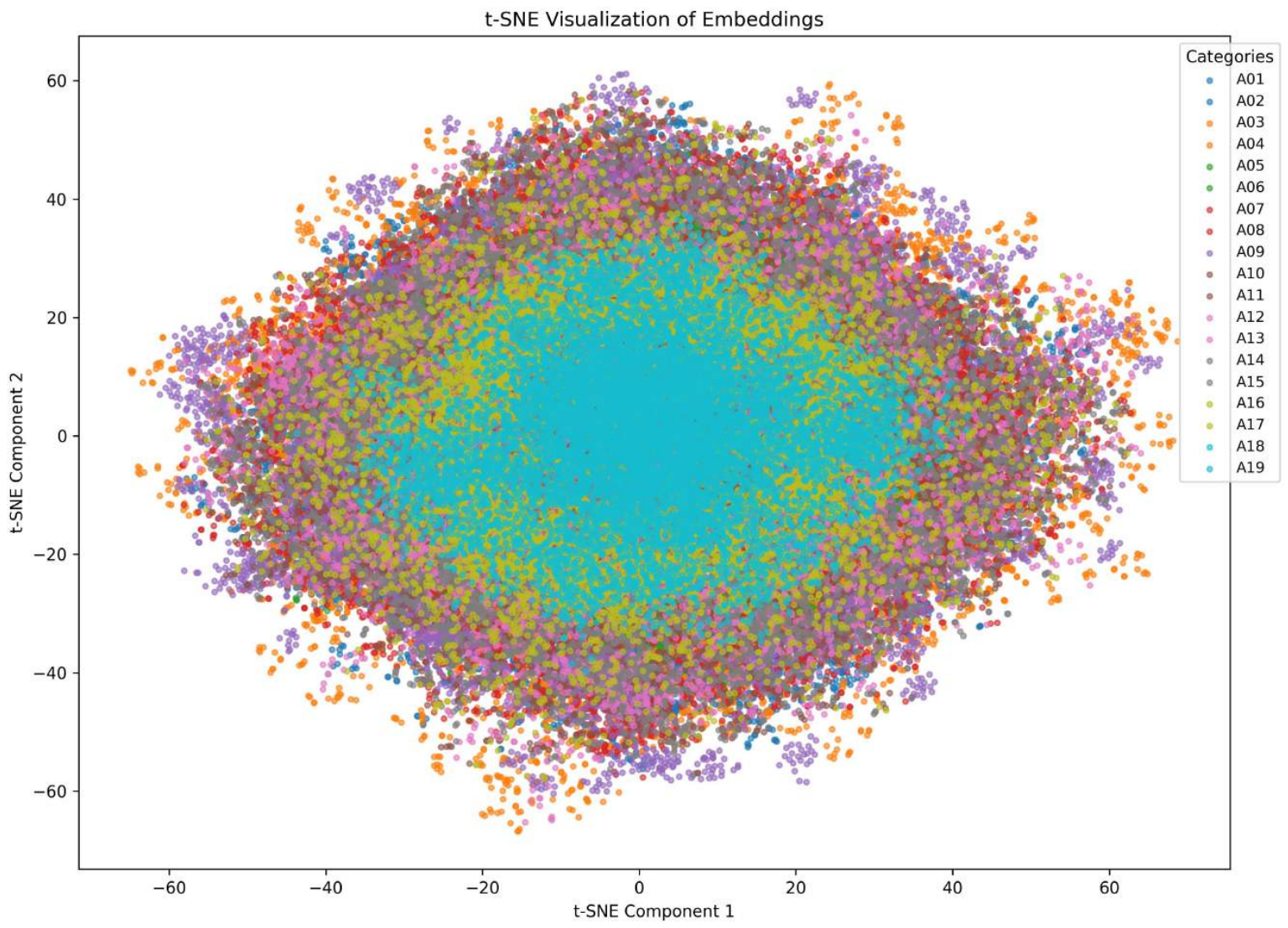}} 
    \subfloat[Wav2vec2-emo]{\includegraphics[width=0.425\textwidth]{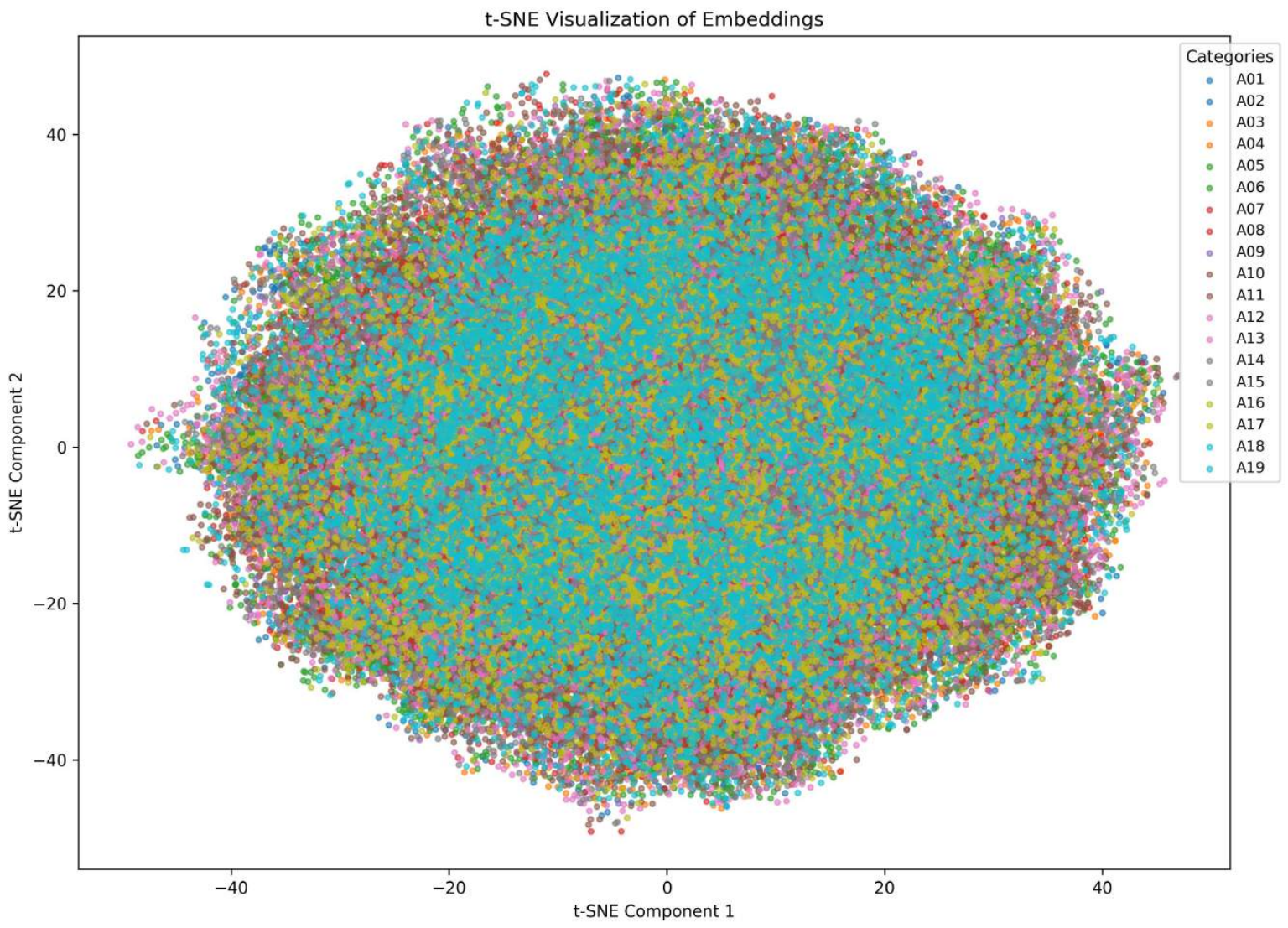}} 

    \caption{Representation Space Visualization of PTMs for ASV}
    \label{tsneasv}
\end{figure*}

\end{document}